\documentclass{osa-article}

\journal{oe}

\articletype{Research Article}

\begin{document}

\title{Transmission properties of microwaves\\
at an optical Weyl point\\
in a three-dimensional chiral photonic crystal
}

\author{
Shun Takahashi,\authormark{1,*} 
Souma Tamaki,\authormark{1} 
Kenichi Yamashita,\authormark{1} 
Takuya Yamaguchi,\authormark{1} 
Tetsuya Ueda,\authormark{1} 
and Satoshi Iwamoto\authormark{2,3}}

\address{
\authormark{1}Kyoto Institute of Technology, Sakyo, Kyoto 606-8585, Japan\\
\authormark{2}Research Center for Advanced Science and Technology, The University of Tokyo, Meguro, Tokyo 153-8505, Japan\\
\authormark{3}Institute of Industrial Science, The University of Tokyo, Meguro, Tokyo 153-8505, Japan
}

\email{\authormark{*}shuntaka@kit.ac.jp} 



\begin{abstract}
Microwave transmission measurements were performed for a three-dimensional (3D) layer-by-layer chiral photonic crystal (PhC), whose photonic band structure contains 3D singular points, Weyl points. 
For the frequency and wavevector in the vicinity of a Weyl point, the transmitted intensity was found to be inversely proportional to the square of the propagation length. 
In addition, the transmitted wave was well-collimated in the plane parallel to the PhC layers, even for point-source incidence. 
When a plane wave was incident on the PhC containing metal scatters, the planar wavefront was reconstructed after the transmission, indicating a cloaking effect. 
\end{abstract}


\section{Introduction}


Singularity in a band structure indicates various exotic phenomena in electrical transport as well as in classical wave propagation. 
One of the famous singularities is the Dirac point in two-dimensional (2D) layered materials such as graphene.
At a Dirac point, two linearly dispersive bands intersect in the 2D momentum space, and ideally induce massless transport, which has been studied for high electron mobility \cite{Neto}.
In addition, because a topological charge appears by lifting a Dirac point degeneracy, topological edge states have been intensively studied not only in electronics \cite{Neto,Armitage} but also in photonics \cite{Ozawa} and phononics \cite{Ma}. 


As a three-dimensional (3D) counterpart of a Dirac point, a Weyl point has also been studied intensively after its experimental discoveries in condensed matter physics \cite{Xu,Lv} and microwave photonics \cite{Lu}. 
To form Weyl points, the necessary condition in the studied systems is to break either spatial inversion ($P$) symmetry or time-reversal ($T$) symmetry \cite{Armitage}. 
Particularly for classical waves that are usually insensitive to external magnetic fields for breaking $T$, most of the experimental reports have adopted to break $P$ in various 3D structures for elastic waves \cite{Nature,CTChan3,Li,YYang}, microwave \cite{Lu,CTChan2,Yang,Science,Jia}, and infrared light \cite{Rechtsman,Gu,Rechtsman2}. 
The use of synthetic dimensions for Weyl points has also been theoretically studied in layered metamaterials \cite{Wang,Yuan}. 
As is the case with Dirac points, topological surface states appear by selecting a particular 2D momentum plane and lifting the Weyl degeneracy apparently \cite{CTChan2,Rechtsman,Yang,Science,Li,Nature,CTChan3,YYang}. 
In four-dimensional band structures, these surface-state modes form open arcs, known as Fermi arcs or surface-state arcs, connecting two Weyl points with opposite signs of topological charges \cite{Armitage,Yang,Li,Nature,YYang}. 
Recent reports show one-dimensional (1D) Hinge states derived from Weyl points \cite{Luo,Wei}. 
In addition, by using dispersion near Weyl points, negative refraction has been reported for elastic waves \cite{Nature,YYang,Yang2021}.


Another exotic phenomenon at such singular points is the characteristic power law of electric current or photonic transmission intensity.  
At a Dirac point, electrical conductivity has been analytically revealed to be constant in both diffusive \cite{Ando} and ballistic \cite{Katsnelson,Beenakker} systems, which is also known as pseudo-diffusive transport. 
Thus, electric current is inversely proportional to the length of conductors. 
In photonic systems, the transmission intensity (photon current) is inversely proportional to the propagation length $L$ \cite{Sepkhanov,Zhang,Dood}. 
In contrast, at the Weyl point, the electrical conductivity has been theoretically found to be $\propto$ $1/L$ in both diffusive \cite{Koshino} and ballistic \cite{Baireuther} systems. 
Analogous to a Dirac point, the transmission intensity at an optical Weyl point may exhibit a power law $\propto$ $1/L^2$, although such propagation properties inside a 3D structure at a Weyl point have not been studied. 
In addition, based on the pseudo-diffusive propagation at Dirac points, phase reconstruction was demonstrated by measuring a planar wavefront of the transmitted field through 2D phononic crystals for a plane-wave injection in the direction parallel to a Dirac-point wavevector \cite{APL,SSC}. 
The zero-refractive-index effects due to accidental Dirac points at the $\it{\Gamma}$ point can also provide wavefront control \cite{CTChan,Kita}. 
However, such characteristic wavefronts in transmitted waves for Weyl points have not yet been investigated. 


In this study, we fabricated all-dielectric 3D chiral photonic crystal (PhC) possessing Weyl points in the photonic band structure, and measured microwave transmission at a frequency of $\sim$ 5 GHz in the vicinity of a Weyl point. 
We found unique transmission properties at the Weyl point, such as a transmission dip at the frequency, and a power law $\propto$ $1/L^2$ of the microwave intensity inside the PhC. 
In addition, the transmitted wave was found to be well-collimated in the plane parallel to the PhC layers, even for point-source incidence. 
We also demonstrated a reconstructed planar wavefront for plane-wave incidence even when the chiral PhC contained metal inclusions, indicating a cloaking effect. 
These results may be significant steps for a deeper understanding of Weyl physics in the future. 


\section{Sample structure}


\begin{figure}
\centering
\includegraphics[width=11cm]{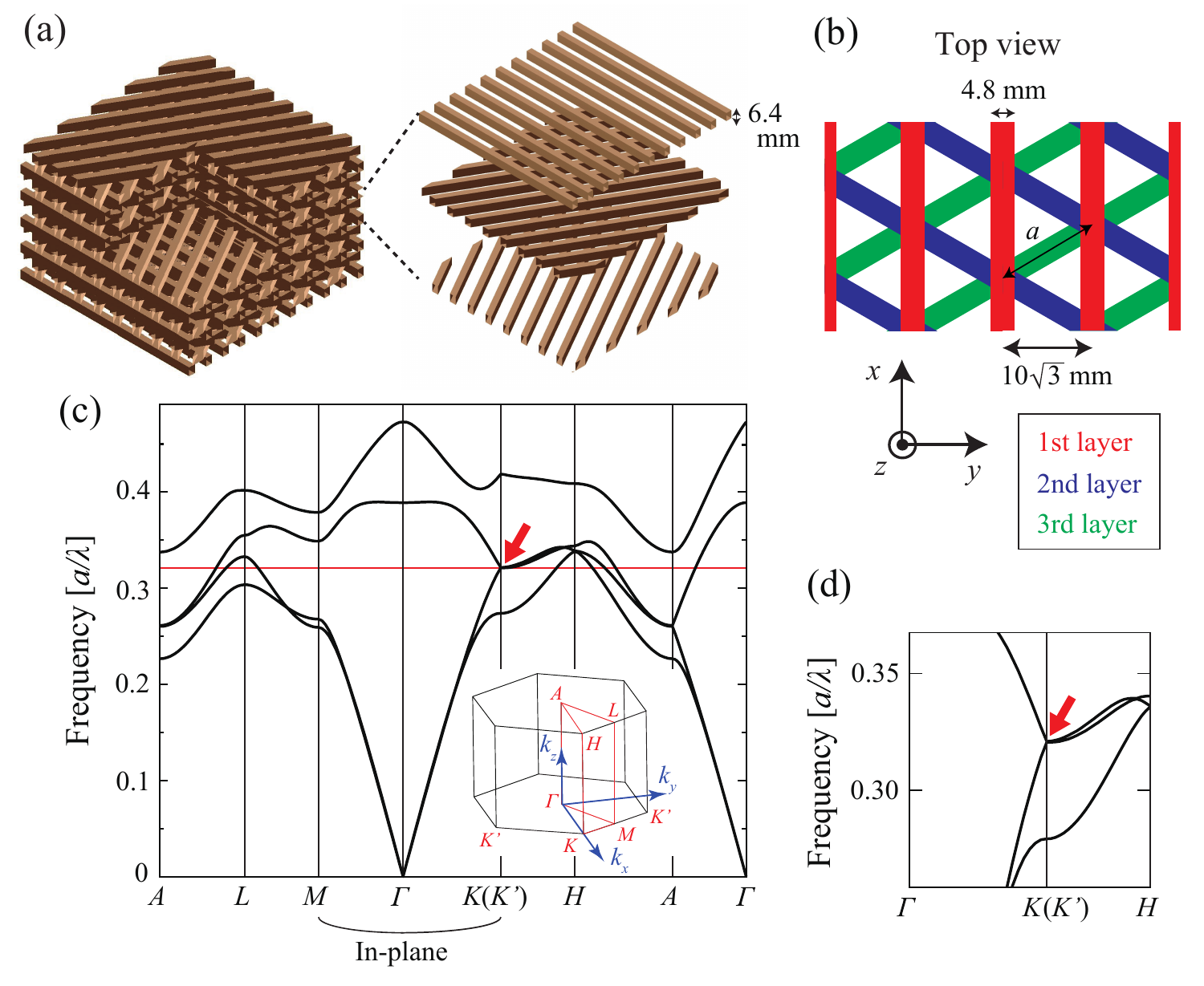}
\caption{
(a)
Schematic of the studied 3D chiral PhC. 
Parts of the structure are removed for clarity. 
(b)
Top view of the chiral PhC. 
To distinguish the stacked layers, the rods for each layer were of different colours.
(c)
Photonic band structure of the chiral PhC for the lowest four bands.
The studied Weyl point at $K$ point is pointed by a red arrow.
The inset shows half of the first Brillouin zone ($k_z$ $>$ 0).
The magnified band structure near the Weyl point is shown in (d). 
}
\label{fig:structure}
\end{figure}


So far, various 3D periodic structures with broken $P$ symmetry have been studied to measure Weyl points. 
Among them, we adopted a chiral woodpile structure in this study (Fig. \ref{fig:structure}(a)). 
In our previous report for this PhC \cite{TakahashiJPSJ}, we calculated the photonic band structure and found 3D point degeneracies. 
In the vicinity of these point degeneracies, we further calculated the section Chern numbers defined for a 2D section of a 3D Brillouin zone. 
These numbers showed non-zero values by changing wavevectors, confirming that the point degeneracies were Weyl points. 
In addition, we proposed the optimum structural parameters of the PhC for practical measurements of the Weyl points. 
Here in this study, to measure one of these Weyl points at microwave operation frequency $\sim$ 5 GHz, the following structural parameters were set by using the scaling law of PhCs \cite{Sakoda}.


The studied layer-by-layer chiral PhC in air is schematically shown in Figs. \ref{fig:structure}(a) and (b).
Each layer was composed of periodically arranged dielectric rods whose refractive indices are $n$ = 3.5 and widths are 4.8 mm.
The rods were arranged in a period of 10$\sqrt{3}$ mm, and the thickness of each layer was 6.4 mm.
These patterned layers were stacked one by one after an in-plane rotation of $60^\circ$ for each stack.
Thus, three layers construct a single helical unit with pitch $p$ = 19.2 mm.
The top view of PhC in Fig. \ref{fig:structure}(b) shows a triangular lattice, and we set an in-plane period $a$ = 20 mm. 


For this chiral PhC, the photonic band structure calculated using the plane-wave expansion method is shown in Fig. \ref{fig:structure}(c) and (d) for the lowest four bands. 
Deduced from the triangular lattice from the top view of PhC, a Dirac-point-like degeneracy occurs at the $K$ and $K'$ points for the second and third bands. 
In contrast to Dirac points in 2D systems, this point degeneracy is lifted from $K$ to $H$ by varying the wavevector $k_z$ parallel to the helical axis, indicating a 3D point degeneracy.
This 3D point degeneracy was confirmed to be a Weyl point by calculating the section Chern number \cite{TakahashiJPSJ}. 
The frequency of this Weyl point is 4.78 GHz for the designed PhC. 

\begin{figure}
\centering
\includegraphics[width=9cm]{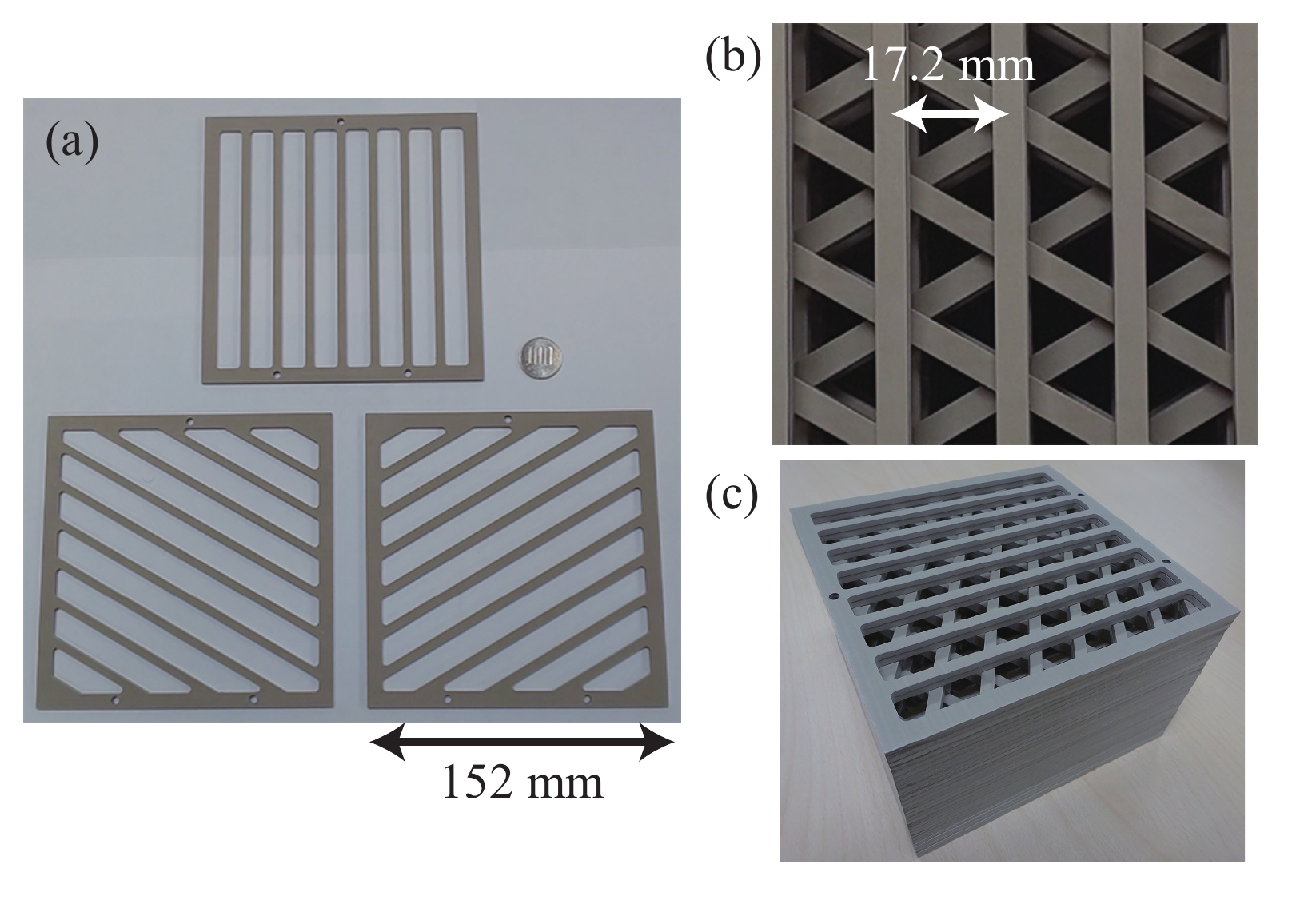}
\caption{
(a) 
Photo of the fabricated three layers with periodic rod patterns.
The rods in each layer are rotated by $60^\circ$.
A 100-yen coin is also placed for comparison.
(b), (c) 
Photo of the stacked 3D PhC from top and bird's eye view, respectively.
}
\label{fig:fab}
\end{figure}

This 3D structure was fabricated by drilling a dielectric board based on polytetrafluoroethylene with $n$ = 3.5 ($\pm$0.1) at $\sim$ 5 GHz. 
The dissipation factor of this material is as small as $\sim$0.002 at approximately 5 GHz.
We carefully selected this dielectric material with $n$ = 3.5, comparable to that of semiconductor materials such as GaAs or Si, for our future applications to telecommunication bands by scaling the 3D PhC to a sub-micron period \cite{TakahashiOE,TakahashiAPL,TakahashiPRB,Si}. 
The dimensions of the fabricated PhC were the same as those of the designed structure, as shown in Fig. \ref{fig:fab}.
The fabrication error was $\pm$0.2 mm. 
We prepared three types of in-plane patterns, where rods were rotated by $60^\circ$ for each pattern. 
The number of in-plane periods is eight in the top pattern in Fig. \ref{fig:fab}(a), and the patterns are surrounded by a self-suspended frame.
By stacking these layers in the appropriate order, we fabricated a 3D chiral PhC with left-handed chirality, as shown in Fig. \ref{fig:fab}(b).
The entire size of the chiral PhC is 152 mm in-plane square and 96 mm height corresponding to five helical periods. 
Owing to the finite diameter of the drill, the rod width near the frame has a relatively large fabrication error of $\sim$3 mm, which is much smaller than the target wavelength of $\sim$60 mm (5 GHz). 
In addition, the frame has three small holes with a diameter of 3.3 mm to align the stacked layers. 


\section{Measurement setups}


For this fabricated PhC, we performed microwave transmission measurements in the $\it{\Gamma}$-$K$ direction using a network analyser (N5224A, Agilent Technologies). 
As an incident source, we first used a loop antenna with a diameter of 3 mm. 
The loop antenna works as a magnetic dipole generating transverse-electric (TE)-polarised waves, where the electric field oscillates in the $x$-$y$ plane parallel to the PhC layers. 
In the phase-reconstruction measurement, we used a horn antenna to generate a plane wave with TE polarisation. 
For the dipole (plane-wave) incidence, the centre of the loop (horn) antenna was positioned 3 (60) mm away from the 2D centre of the PhC side surface of incidence. 


For spatially resolved detection of the TE-polarised microwave, another loop antenna was mechanically scanned in a 3D spatial resolution of 2.5 mm. 
For the intensity measurement inside the PhC, the detection antenna was inserted into the triangular holes. 
From the network analyser, we obtained one of the scattering parameters, $S_{21}$, between the source loop/horn antenna (port 1) and the detection loop antenna (port 2) in the frequency range of 4 - 6 GHz with the resolution of 0.04 GHz. 
We then extracted amplitude and phase of the transmitted microwave at each frequency. 
During the measurements inside (outside) the PhC, we attached absorber sheets, 15 dB absorption at approximately 5 GHz, on the surfaces of the PhC, except for the incidence and top (transmission) surfaces. 


\section{Transmission intensity at the Weyl point}


\begin{figure}
\centering
\includegraphics[width=11cm]{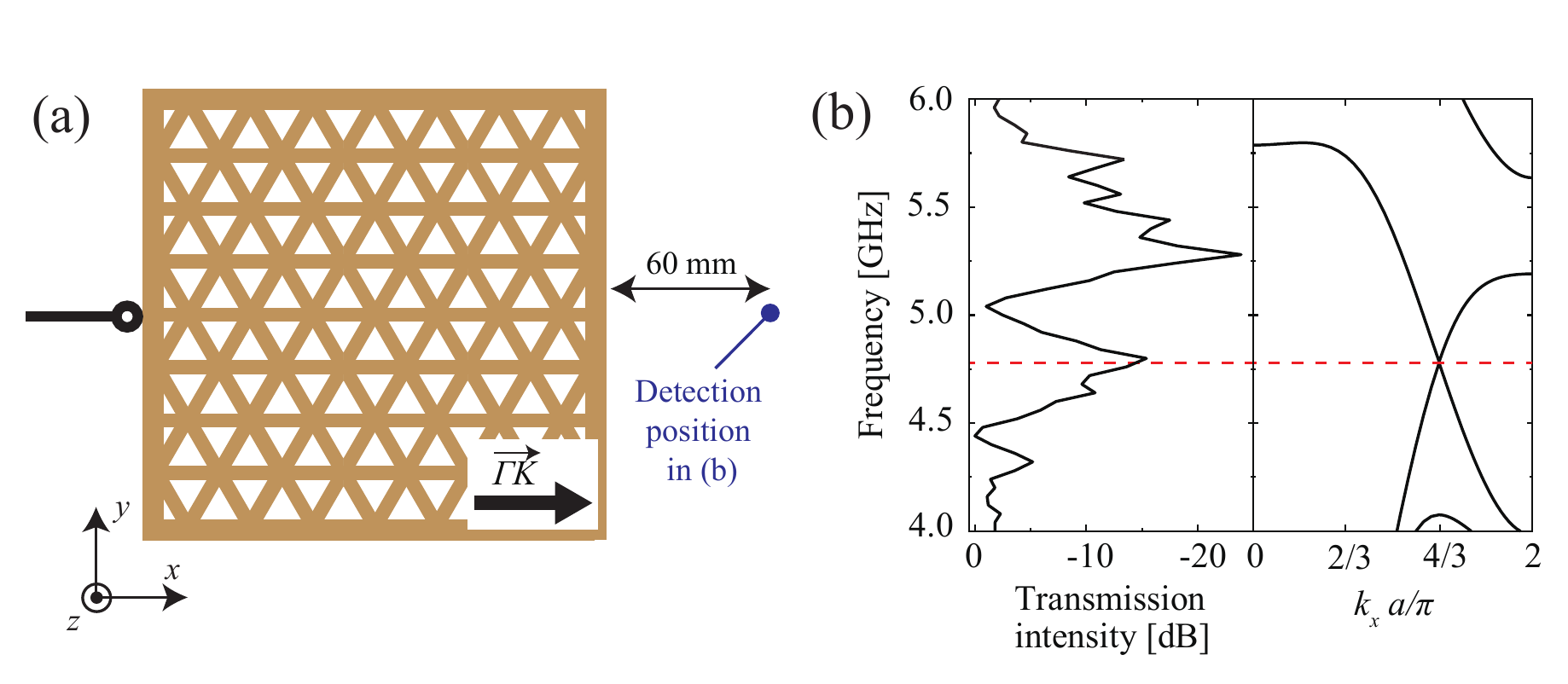}
\caption{
(a)
Schematic of the measurement for the transmission spectrum. 
(b)
Measured transmission spectrum (left panel) and corresponding band structure in the $k_x$ direction (right panel).
}
\label{fig:spectrum}
\end{figure}


First, we measured a transmission spectrum for the dipole incidence. 
The transmitted intensity was measured from 60 mm (single wavelength) away from the backside of the PhC, as schematically shown in Fig. \ref{fig:spectrum}(a). 
Figure \ref{fig:spectrum}(b) shows the transmission spectrum in the left panel, accompanied with the calculated band structure in the $k_x$ direction in the right panel. 
The details of the band calculation are shown in \textcolor{urlblue}{Supplement 1}. 
In the transmission spectrum, the intensity was decreased by 15 dB at approximately 4.75 GHz, corresponding to the Weyl point frequency in the right panel. 
Because the density of states at an ideal Weyl point is zero \cite{Koshino,GarciaElcano}, this intensity decrease experimentally indicates the existence of the Weyl point. 
In the high-frequency region at approximately 5.50 GHz, only a small transmission intensity was detected because of the symmetry mismatch in the spatial distribution of the electromagnetic field between the TE incidence and the third lowest band for $k_x a$ < $4\pi/3$ (shown in \textcolor{urlblue}{Supplement 1}). 
Similar spectra were obtained at different detection positions for 30 mm < $x$ < 120 mm away from the PhC surface. 
From this agreement between the experimental spectrum and the numerical band structure, we confirmed that the Weyl point exists at approximately 4.75 GHz. 

\begin{figure}
\centering
\includegraphics[width=11cm]{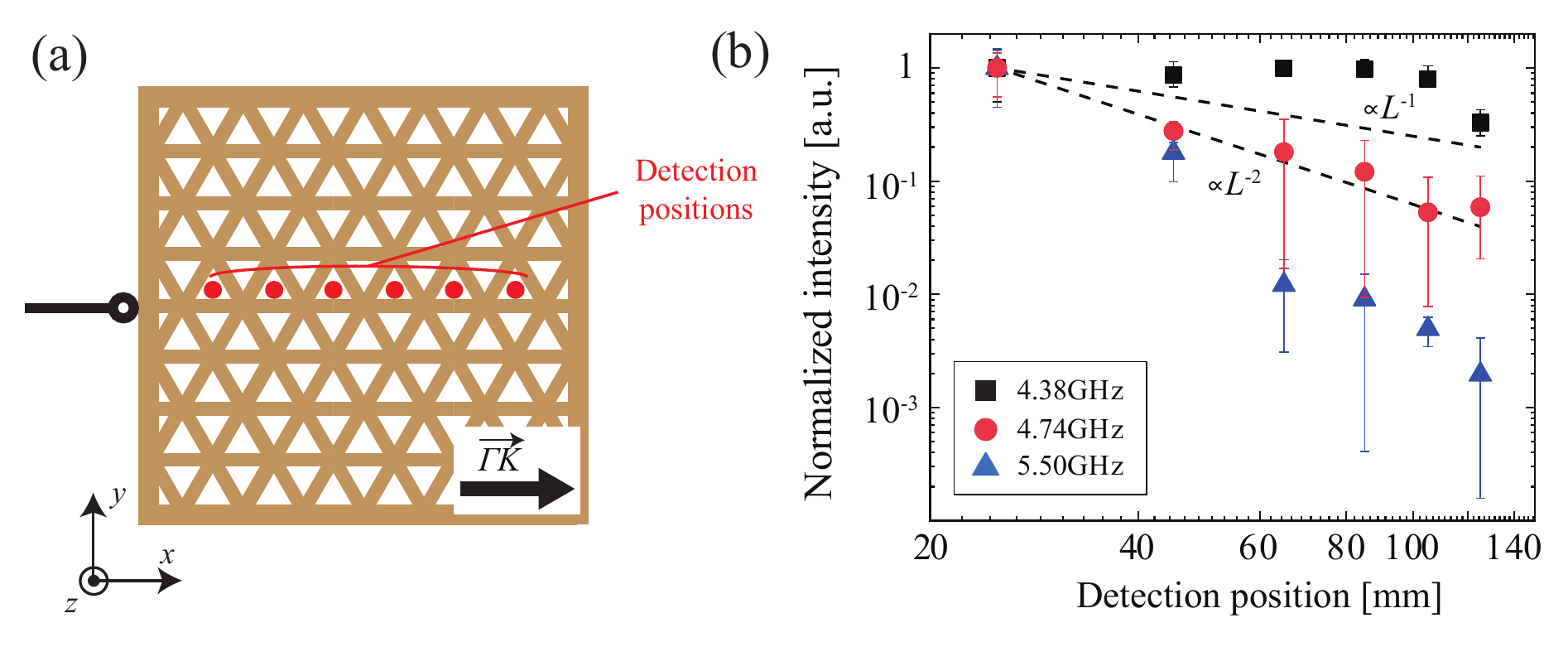}
\caption{
(a)
Schematic of the intensity measurement for the detection position dependence inside the PhC. 
(b)
Microwave intensity inside the PhC as a function of distance from the source, $L$.
The detected intensity was normalised to the value at $L$ = 25 mm for each frequency.
The two dotted lines indicate the power-law of $L^{-1}$ and $L^{-2}$. 
}
\label{fig:intensity}
\end{figure}

Thereafter, we measured the microwave intensity inside the PhC by inserting the detection antenna into the six triangular holes, as schematically shown in Fig. \ref{fig:intensity}(a).
The height of the detection antenna was scanned from -5 to +5 mm relative to the height of the source. 
Figure \ref{fig:intensity}(b) shows the detected intensity as a function of the distance from the source, $L$, for three particular frequencies; 4.38 GHz, 4.74 GHz, and 5.50 GHz, corresponding to the symmetry-matched band, Weyl point, and symmetry-mismatched (anti-symmetric) band, respectively. 
The symbols (error bars) indicate the average (distributed) intensity for each height of the detection antenna. 
The measured intensity was normalised to the value at $L$ = 25 mm for each frequency. 
At 4.38 GHz, the intensity was large and almost independent of $L$ owing to the symmetry-matched band. 
At 5.50 GHz, the intensity decreased exponentially because the symmetric input field could not efficiently excite the anti-symmetric Bloch mode. 
In contrast, at 4.74 GHz, the intensity was inversely proportional to the square of $L$, which coincides with the theoretical results for ballistic electronic systems \cite{Baireuther}. 
This power law is clearly different from that for Dirac points in 2D photonic systems where the transmission intensity $\propto$ $1/L$ \cite{Sepkhanov,Zhang}, implying that the transport at the Weyl point was not pseudo-diffusive. 
Such difference in the power law stems from the dispersion of the density of states in the vicinity of a Dirac/Weyl point \cite{Koshino}. 
The density of states shows a linear dispersion for a Dirac point, whereas that shows a quadratic dispersion for a Weyl point. 
Numerical calculations also showed the power law $\propto$ $1/L^2$ (shown in \textcolor{urlblue}{Supplement 1}). 
These experimental and numerical results demonstrate a unique aspect of optical Weyl points, and a rigorous theoretical analysis of a photonic system is expected. 


\section{Spatial distribution of the transmitted wave}


\begin{figure}
\centering
\includegraphics[width=13cm]{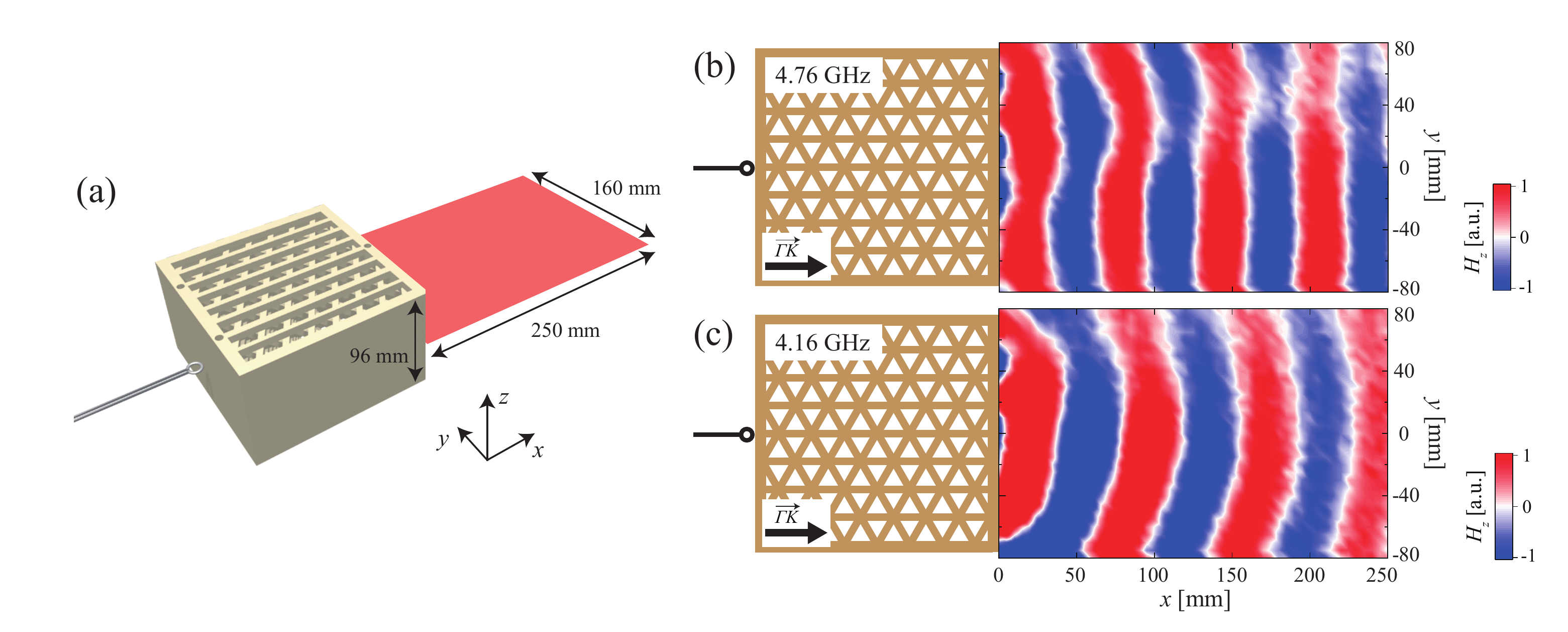}
\caption{
(a)
Schematic of the $x$-$y$ detection plane at the backside of the PhC. 
(b), (c)
Spatial distribution of the transmitted wave at 4.76 GHz and 4.16 GHz, respectively.
}
\label{fig:xy_plot}
\end{figure}


\begin{figure}
\centering
\includegraphics[width=11cm]{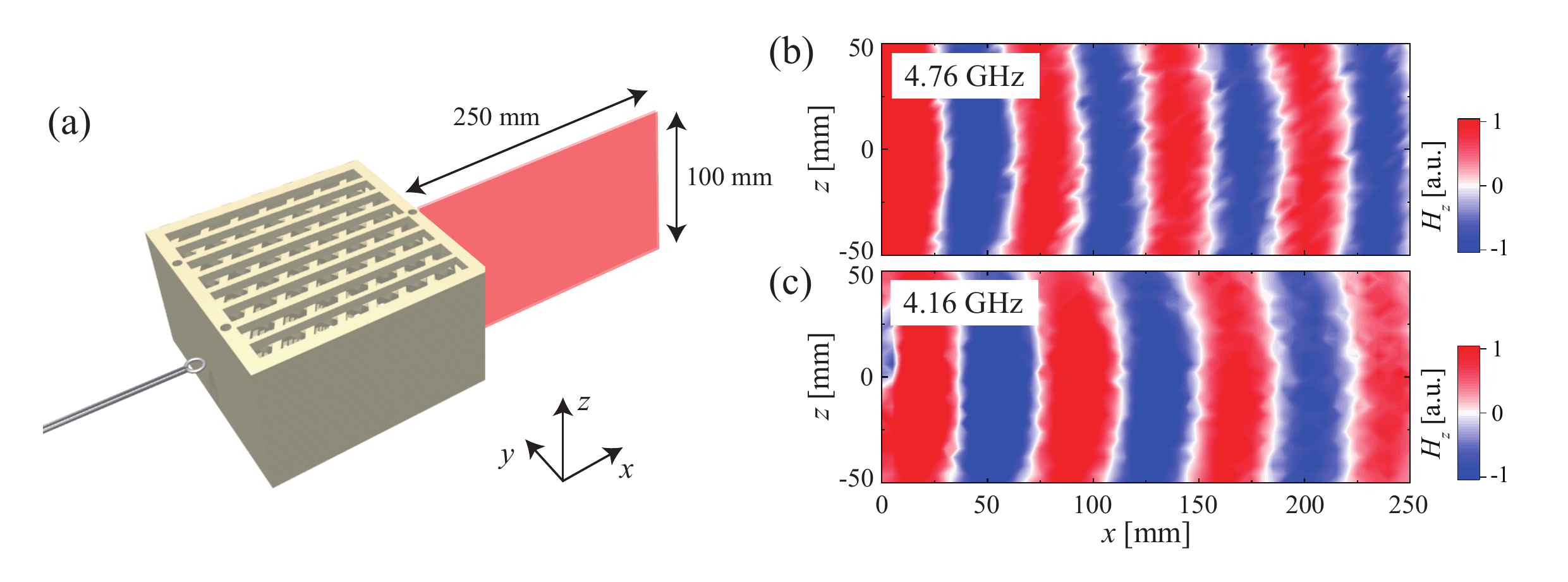}
\caption{
(a)
Schematic of the $x$-$z$ detection plane. 
(b), (c)
Spatial distribution of the transmitted wave at 4.76 GHz and 4.16 GHz, respectively.
}
\label{fig:xz_plot}
\end{figure}


We investigated spatial distribution of the transmitted wave at the backside of the PhC. 
We first measured both the amplitude and phase of the transmitted microwave in the $x$-$y$ plane at the same height as the dipole source, as schematically shown in Fig. \ref{fig:xy_plot}(a).
Figures \ref{fig:xy_plot}(b) and (c) show the spatial distribution of the transmitted wave at the frequency of the Weyl point (4.76 GHz) and in the continuous band (4.16 GHz), respectively. 
Whereas a spherical wavefront was detected in (c), a planar wavefront was observed in (b), even for the point-source incidence in the 2D plane. 
Note that such an output with a planar wavefront was never observed in the $\it{\Gamma}$-$M$ direction at 4.76 GHz (shown in \textcolor{urlblue}{Supplement 1}), nor at different frequencies from the Weyl point. 


We also measured the spatial distribution of the transmitted wave in the $x$-$z$ plane at the same $y$ position as the dipole source, as schematically shown in Fig. \ref{fig:xz_plot}(a).
Figures \ref{fig:xz_plot}(b) and (c) shows the mapped transmitted waves for the $y$ slice at 4.76 GHz and 4.16 GHz, respectively. 
Fig. \ref{fig:xz_plot}(c) at 4.16 GHz again shows a clear spherical wavefront, whereas Fig. \ref{fig:xz_plot}(b) at the Weyl point frequency does not show a clear planar wavefront, but shows a spherical wavefront. 
From these results, the output from the chiral PhC at the Weyl point frequency was found to have a cylindrical wavefront. 

\begin{figure}
\centering
\includegraphics[width=10cm]{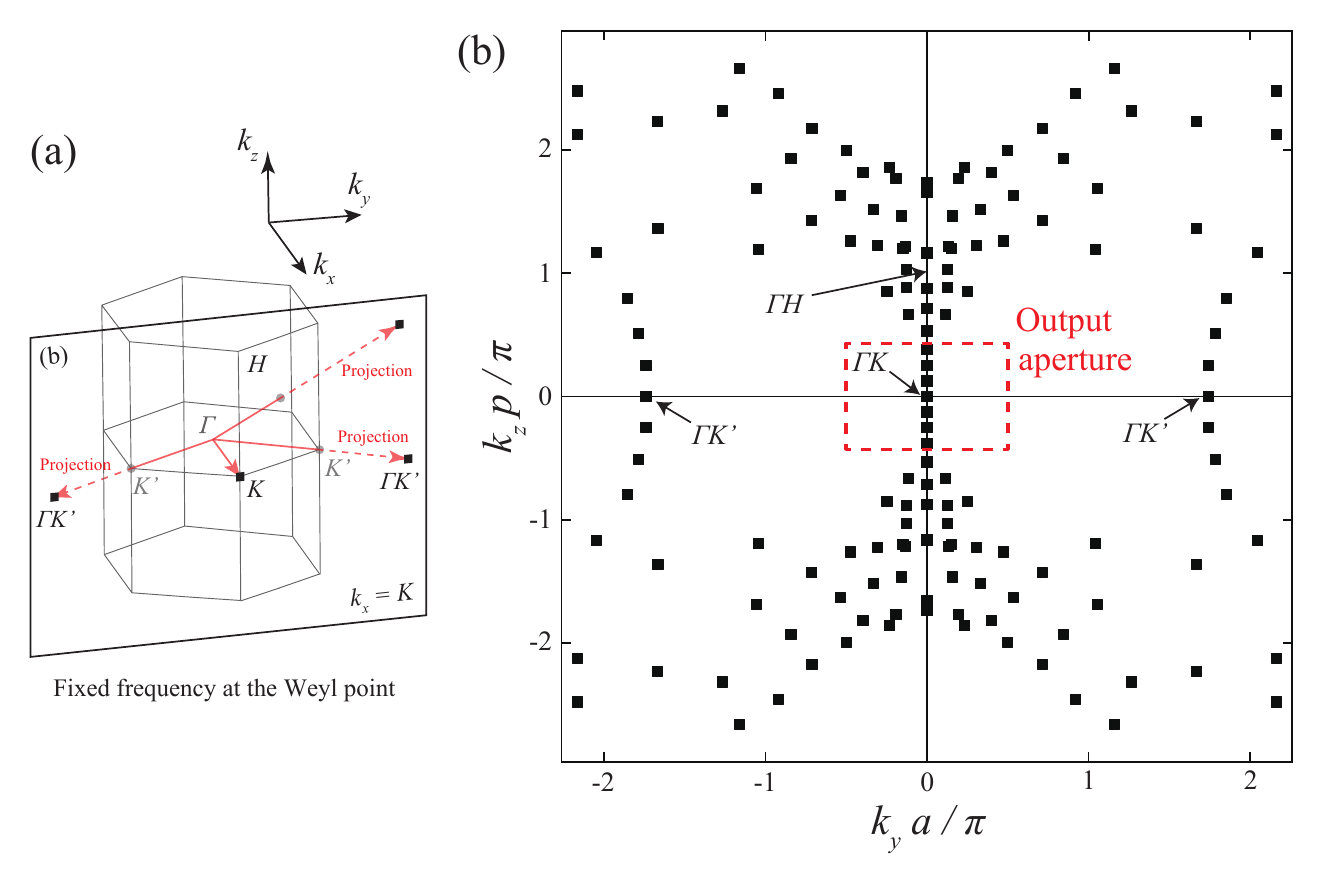}
\caption{
(a)
Schematic of the projection on the $k_x$ = $K$ plane from $\it{\Gamma}$ for the Bloch modes at the Weyl point frequency. 
(b)
Projection result of (a). 
In the red dotted square corresponding to the output aperture of the PhC, the projected Bloch modes exist only on the $k_z$ axis continuously. 
}
\label{fig:projection}
\end{figure}

To explain these output wavefronts for the dipole input, we first consider a $k$-space distribution of the Bloch modes in the PhC. 
As shown schematically in Fig. \ref{fig:projection}(a), we numerically projected the Bloch modes at the Weyl point frequency in the first Brillouin zone onto the $k_x$ = $K$ plane from $\it{\Gamma}$ point. 
Figure \ref{fig:projection}(b) shows the projection result with discrete plots because the first Brillouin zone was divided into $16^3$ meshes in our calculation of the equi-frequency dispersion curves. 
Thereafter, we calculated the angles between $\overrightarrow{\it{\Gamma}K}$ and the vectors from the dipole source to the sides of the output surface.
For the $k$ vectors corresponding to these angles, the collection of their components ($k_y$, $k_z$) at $k_x$ = $K$ represents the red dotted square in Fig. \ref{fig:projection}(b). 
Here, we refer to this square as the output aperture, and the projected modes outside this square are all absorbed by the absorber sheets attached to the PhC. 


Based on the numerical results shown in Fig. \ref{fig:projection}(b), we first discuss the propagation in the in-plane directions at $k_z$ = 0.
In the output aperture, there is only a single mode of the Weyl point. 
Because waves from the dipole source can be represented by a sum of Fourier series of plane waves with various wavevectors, one of the Fourier components whose wavevector is same as that of the Weyl point can transmit through the PhC. 
Therefore, the PhC filtered the single plane wave and the planar wavefront was obtained, as shown in Fig. \ref{fig:xy_plot}(b). 
Further numerical calculations also yielded a well-collimated output (shown in \textcolor{urlblue}{Supplement 1}).
In contrast, in the out-of-plane propagation, there are continuous modes along the $k_z$ axis in the output aperture in Fig. \ref{fig:projection}(b).
Thus, the transmitted wave showed a spherical wavefront. 
In the case of the ideal Weyl point \cite{Science}, the transmitted wave could exhibit a planar wavefront in both the in-plane and out-of-plane directions. 


\begin{figure}
\centering
\includegraphics[width=9cm]{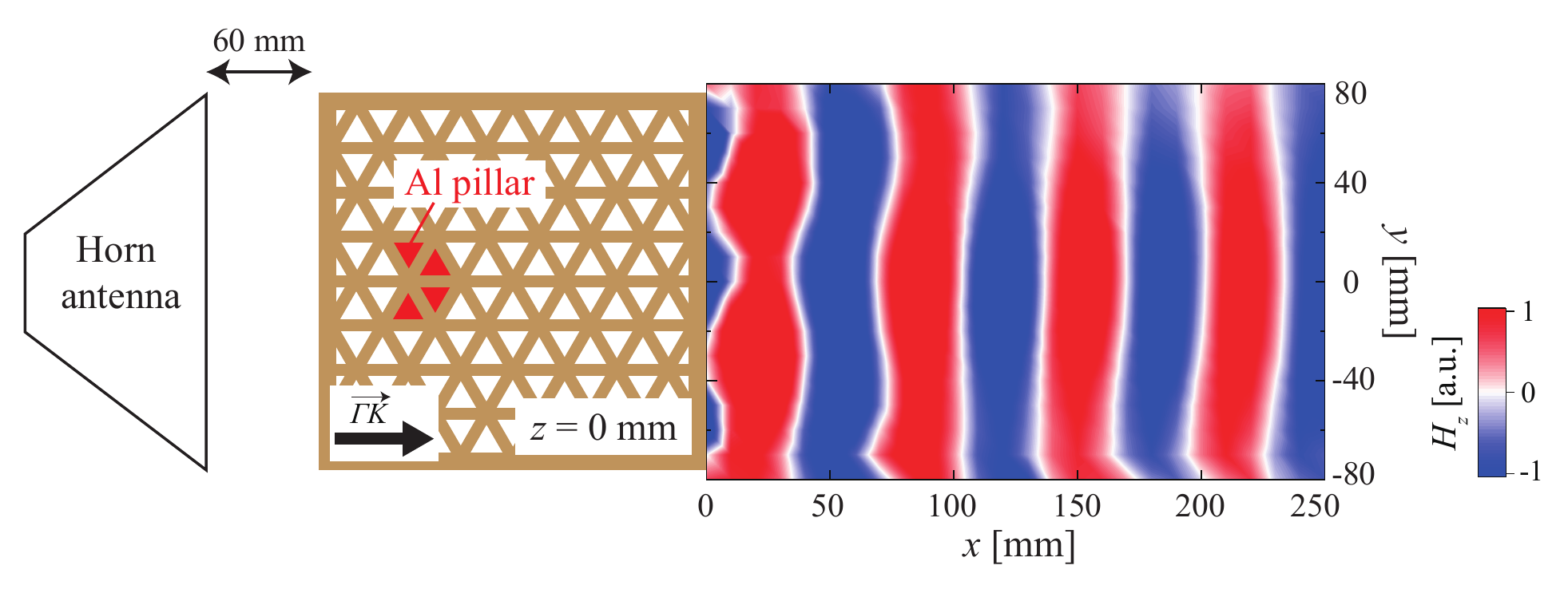}
\caption{
Spatial distribution of the transmitted wave for the PhC containing four Al pillars.
The input plane wave at 4.76 GHz was generated by the horn antenna, and the detection antenna was scanned in the $x$-$y$ plane at the central height of the PhC.
}
\label{fig:cloaking}
\end{figure}

By utilising the well-collimated output for the point-source input in the $x$-$y$ plane at the Weyl point frequency, we demonstrated a cloaking effect in the 2D plane, where phases of plane-wave incidence are reconstructed after transmission through a scattering material. 
We intentionally inserted four regular triangular pillars made of aluminium into the holes of the PhC, as shown in Fig. \ref{fig:cloaking}. 
Each side of the triangle and the length of the pillars are 11.7 mm and 96 mm, respectively, which are the same dimensions as the triangular holes of the PhC. 
For this PhC containing these four scatterers, we applied a plane wave in the $\it{\Gamma}$-$K$ direction by using the horn antenna, and measured the spatial distribution of the transmitted wave at the backside of the PhC. 
Figure \ref{fig:cloaking} shows a planar wavefront output despite the existence of the scatterers, indicating a phase-reconstruction effect. 
This result could be interpreted as a cloaking effect in the direction parallel to that of the Weyl-point wavevector. 


\section{Summary}


We investigated the microwave transmission near the frequency and wavevector of the Weyl point in the all-dielectric 3D chiral PhC. 
The intensity measurement inside the PhC exhibited the spatial power law $\propto 1/L^2$, which coincides with the previous theoretical report on electronic systems \cite{Baireuther}.
After transmission through the PhC, we observed an output with a planar wavefront even for the point-source input in the plane parallel to the PhC layers, because the PhC filtered the single plane-wave mode of the Weyl point. 
This 2D well-collimated output was not disturbed by metal inclusions in the PhC, which could be applied to the cloaking effect. 
These results may provide important insights for the future Weyl physics and applications to control microwave propagation such as the cloaking effect, a Veselago lens \cite{Yang2021} based on the negative refraction \cite{Nature}, and topologically protected propagation channels including 1D Hinge states \cite{Luo,Wei}. 


\section*{Funding}
This work was supported by Grants-in-Aid for Scientific Research, Nos. 17H06138, 15H05700, 18K18857, and JST-CREST JPMJCR19T1.


\section*{Acknowledgments}
The authors would like to thank M. Koshino for his fruitful discussions. 


\section*{Disclosures}
The authors declare no conflicts of interest.


\section*{Data availability}
Data underlying the results presented in this paper are not publicly available at this time but may be obtained from the authors upon reasonable request. 


\section*{}
See \textcolor{urlblue}{Supplement 1} for supporting content.



\end{document}